# ODE Network Model for Nonlinear and Complex Agricultural Nutrient Solution System

Byunghyun Ban*
Andong District Office
**Ministry of Employment and Labor**
Andong, Republic of Korea
bhban@kaist.ac.kr

Minwoo Lee
Future Agriculture Team
**Imagination Garden Inc.**
Andong, Republic of Korea
hydrominus@sangsang.farm

Donghun Ryu
Machine Learning Team
**Imagination Garden Inc.**
Andong, Republic of Korea
dhryu@sangsang.farm

*Abstract*—In closed hydroponic systems, periodic readjustment of nutrient solution is necessary to continuously provide stable environment to plant roots because the interaction between plant and nutrient solution changes the rate of ions in it. The traditional method is to repeat supplying small amount of premade concentrated nutrient solution, measuring total electric conductivity and pH of the tank only. As it cannot control the collapse of ion rates, recent researches try to measure the concentration of individual components to provide insufficient ions only. However, those approaches use titration-like heuristic approaches, which repeat adding small amount of components and measuring ion density a lot of times for a single control input. Both traditional and recent methods are not only time-consuming, but also cannot predict chemical reactions related with control inputs because the nutrient solution is a nonlinear complex system, including many precipitation reactions and complicated interactions. We present a continuous network model of the nutrient solution system, whose reactions are described as differential equations. The model predicts molar concentration of each chemical components and total dissolved solids with low error. This model also can calculate the amount of chemical compounds needed to produce a desired nutrient solution, by reverse calculation from dissolved ion concentrations.

## I. Introduction

Recently, soilless culture takes center stage in agricultural industry. Closed hydroponic system is one of the most popular hydroponic method because it reduces the cost and hazard of water pollution [1]. As plants continuously absorb nutrients from the environment, the concentration of individual ions continuously drops. Traditional methods usually measure pH and electrical conductivity (EC) of the nutrient solution to monitor the fertilization status [2-3]. When EC is low, they add premade concentrated solution to the tank and then apply acids to maintain pH.

As the absorption rates of the ions are all different, those approaches gradually destroy the ratio among ions [4] and accumulates excessive ions (sodium, chloride, sulfate and etc.) [5-6] which have low absorption rates or are supplied too much. Many researchers recently have suggested to measure individual ion with ion-selective sensors and to provide insufficient ions only [1, 7-9]. However, their control methods are slow and cannot avoid Na+ accumulation problem caused by Fe-EDTA supply.

Nutrient solution is a complex system. It is a bi-directed network model, whose nodes are chemical components and edges are reactions. It is difficult to figure out the exact state, and some input can cause unexpected results because almost all the vesicles have self-feedback structures or bi-directed interactions. And many reactions lead to undesired output nodes such as sediment or unabsorbable ions. For example, supplying additional chemicals does not just raise ion concentrations directly. The components in the nutrient solution make various reactions such as sedimentations or reductions, producing compounds which plant does not absorb. As researchers does not know what is happening in the nutrient solution system exactly, they proposed some models to predict salt accumulation [6] or ion rates [4].

Boolean network model and ordinary differential equation (ODE) model are frequently applied to describe complex system. Boolean system describes value of the components as true or false binary. So construction of large-scale network model such as cancer cell model [10-13] is a novel and useful approach. ODE network describes interaction between components as ordinary differential equations, which usually have time t as independent variable [14-16]. It requires huge computing power, and it is difficult to build differential equations for the whole network. ODE model can describe continuous system while Boolean model can describe discrete phenomena only. Applying Boolean network on nutrient system modeling can only show existence of a component as true or false value but ODE network can describe continuous changes of concentrations of ions and sedimentation reactions.

Chemical reactions are time-dependent continuous process so they can be modeled as ordinary differential equations, whose independent variable is the time. For example, a sedimentation reaction in the nutrient solution $CaSO_4 \rightleftharpoons Ca^{2+} + SO_4^{2-}$ is described as equation (1). The coefficients $k_1$ and $k_2$ are reaction rate coefficients which shows how fast the reaction is. If one component appears on the left side of various differential equations, they can be superpositioned as single equation. As a chemical reaction influences every component except catalyst, chemical reaction network has a lot of self-feedbacks. If a chemical produces more same ions at the same time, we multiplied the number of ions on the reaction rate coefficient like equation (2) describing $Ca(NO_3)_2$ dissociation, in terms of $NO_3$.

$$\frac{d[CaSO_4]}{dt} = k_1[Ca^{2+}][SO_4^{2-}] - k_2[CaSO_4]$$

$$\frac{d[Ca^{2+}]}{dt} = -k_1[Ca^{2+}][SO_4^{2-}] + k_2[CaSO_4]$$
$$\frac{d[SO_4^{2-}]}{dt} = -k_1[Ca^{2+}][SO_4^{2-}] + k_2[CaSO_4] \quad (1)$$

$$\frac{d[NO_3^-]}{dt} = 2k_1[Ca(NO_3)_2] - k_2[Ca^{2+}][NO_3^-][NO_3^-] \quad (2)$$

Although the topology of chemical system network is easily driven from known reaction sets, the reaction rate coefficient is not. It is measured by experiments [17]. The coefficient of each equation defines response of the system because kinetic parameters define the activity of the equation. However, precise literature values for those chemical processes in nutrient solution system are missing. Parameter estimation algorithms for complex network systems have been proposed in systems biology field [18-20] but they require experimental data. It is not feasible to measure the amount of all the chemical compounds in nutrient solution along time adding input, because not all kinds of ion selective electrodes (ISEs) and sediments are not measurable with commercially available sensors, while not affecting any chemical environments such as pH or temperatures.

We present a comprehensive and persuasive network model for nutrient solution system whose parameters are driven from literature values. The kinetic parameters are based on equilibrium constants. This model can simulate both forward and reverse reaction at the same time, and even can perform time-reverse simulation. Simulation with this model is easily perform without GPU devices.

We can predict the ionic composition and the amount of sediments by dissolution simulation of fertilization materials. Even calculation of the amount of each nutrient powder from ionic solution state is possible with reverse-direction simulation. As it is a white-box model, it can also trace the accumulation of $Na^+$ or other unabsorbable ions in closed hydroponic system. Readjustment method for nutrient solution should also be changed because the model can show the amount of required materials. Pouring a shot of chemicals into the tank is enough, rather than traditional methods which take several minutes for single step of control input.

## II. METHODS

### A. Network Topology Design

We established nutrient solution system with Yamazaki's solution for Lactuca Sativa L. [21], which includes N, P, K families and microelements in highly-plant-absorbable ion state. Although the industrial recipes recommend hydrates [22], there are too many possible numbers of water molecules per formula unit for one salt, and even incomplete sealing increases it during storage in fields. So we chose dehydrated chemical compounds in order to build a standard model for nutrient solutions.

The selected standard chemicals, their ionized forms and the products which are produced by reactions among the ions which are involved the experiment of the experiments are provided on S1 in the supplementary information section. They are all enlisted on the system network. We also added water, hydrogen ion, hydroxyl ion, nitric acid for pH adjustment and UV light which disintegrates Fe-EDTA$^-$ ion to make the model more comprehensive.

The interaction among nodes are simply classified into 3 classes: enhancement, suppression, not-interactive. Fig 1 is visualized network topology with Cytoscape [23].

Dissolution of system input is regarded as irreversible processes because nutrient solution is thin enough and external plant interference which reduces concentrations of plant-absorbable ions is applied in real world.

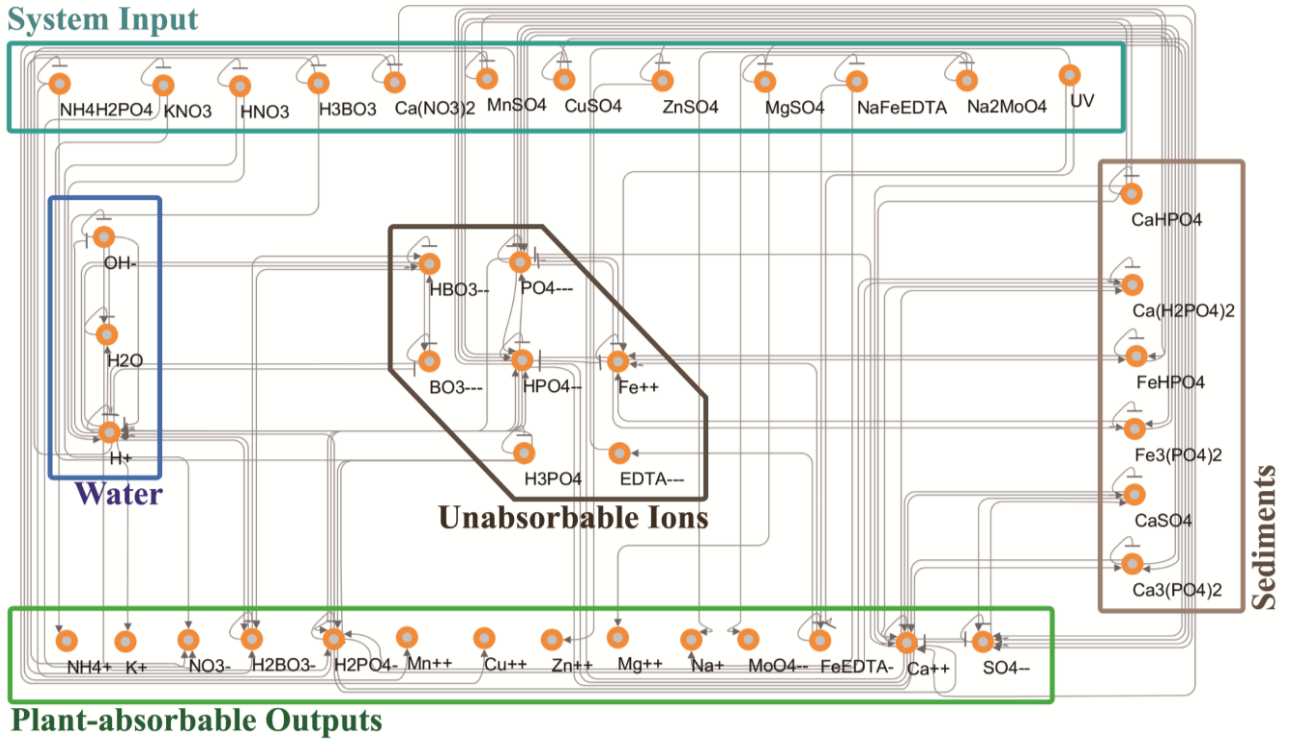

**Fig. 1. Nutrient Solution System Network Model.**

*B. Network Dynamics Design*

Reaction rate coefficients are driven from equilibrium constants by the relation $K = k_f/k_b$. Equilibrium constant is a description of a state of convergence so if we manipulate the value of $k_f$ and $k_b$ while fixing their ration as K, the reaction converges within the same status if time stride is not too big. Therefore, if we let $k_f = K$ and $k_b = 1$, the differential equation converges appropriately because their rate is still K, but the required time to make convergence only differs. It also can describe ionization and its reverse process, which are not actually chemical reactions.

We used dissociation constant $K_d$, which explains a reversible process in which big components breaks down into smaller components, to explain liquid state input such as nitric acid. For example, the equilibrium state of a reversible dissociation process $X_a Y_b \leftrightarrows aX + bY$ is described as equation (3). $K_d$ is directly obtain by gathering literature value of acid dissociation constant $K_a$. The solubility product equilibrium constant $K_{sp}$ is adopted for dissolution of solids because it describes the dynamic equilibrium between solid and ion state. Solubility product equilibrium constant for dissolution $Z_{(s)} \leftrightarrows aX_{(aq)} + bY_{(aq)}$ is defined as equation (4), which is easily calculated with solubility.

$$K_d = \frac{[X]^a [Y]^b}{[X_a Y_b]} \quad (3)$$
$$K_{sp} = [X_{(aq)}]^a [Y_{(aq)}]^b \quad (4)$$

Although simple application of $k_f$ and $k_b$ driven above can predict converged state of the system, it is not enough to describe system states before convergence. Therefore, we multiplied correction coefficient, which is a positive number between 1 and 1000, to correct the speed of forward and backward reactions. For example, dissociation of $HNO_3$ into $H^+$ and $NO_3^-$ is very fast but dissolution of NaFe-EDTA into $Na^+$ and Fe-EDTA$^-$ is relatively slow. So we multiplied 1000 to both $k_f$ and $k_b$ for nitric acid dissolution process and multiplied 1 to those of ferric EDTA.

Literature values of $K_a$, solubility for all reactions and calculated equilibrium constants involved in the experiments are provided on S2 in the supplementary information section. The differential equations, their reaction rate coefficients and correction coefficients used to perform simulations are listed in Table S3 on the supplementary information section.

III. ALGORITHM

The system needs initial state information. As the kinetics are time-dependent, the model is an ordinary differential equation model. An ODE model updates its next-step status by applying current-state information. Let $[X_i]$ the concentration of i-th node variable of the network. It is a time-dependent variable. In other word, $[X_i]$ is a function of time. ODE for $[X_i]$ is equation (5), where q is total number of terms in superpositioned equation for $[X_i]$. Coefficient c means the number of $X_i$ in the term.

$$\frac{d[X_i](t)}{dt} = \sum_{j=1}^{q} \{ck_j \prod [input\ to\ X_i]\ (t-1)\} \quad (5)$$

$$[X_i](t) = [X_i](t-1) + \frac{d[X_i](t)}{dt} \Delta t \quad (6)$$

Update for $[X_i]$ along time is performed with gradient-descent-like method as equation (6, where $\Delta t$ is the time interval. If the time interval is too small, the system converges too slowly but if it is too big, the system may not converge.

We applied synchronous update method, which update all $[X](t)$ values from all $[X](t-1)$, because chemical reactions in one solution occurs simultaneously. We defined X(t) as a vector of concentration of the nodes and simply calculated $\frac{dX(t)}{dt}$ to obtain gradient vector. Synchronous update with vector-form is described in equation (7).

$$X(t) = X(t-1) + \frac{dX(t)}{dt} \Delta t \quad (7)$$

We built a chemical complex system solver with Python language. This simulator receives a text file containing differential equations of chemical reactions and reaction rate coefficient and automatically performs superposition for overlapped variables. The chemical topology is driven during text parsing process. It also performs both forward and backward simulations. We also wrote a text file which contains the topological and kinetic information of nutrient solution model. Both chemical complex system solver and nutrient system model information file are provided as an open sourced python package at the authors' Github repository: https://github.com/needleworm/nutrient_solution.

IV. EXPERIMENT

To examine the performance of the simulator, we performed experiment on Yamazaki's nutrient solution. As ISEs have error in complex chemical system due to the ion interference phenomenon, traditional methods to selectively readjust individual ions are not feasible. We compared ISE-observed ion concentration with the model's prediction, as well as the literacy value.

*A. Equipments*

Vernier's Go Direct® ISE series, GDX-NO3, GDX-NH4, GDX-CA, GDX-K, are used to measure the concentration of ions. $KNO_3$, $Ca(NO_3)_2 \cdot 4H_2O$, $NH_4H_2PO_4$, $MgSO_4 \cdot 7H_2O$ are used to produce Yamazaki's nutrient solution for lettuce. The simulation was performed on Intel's I7-6850K with Python 3.6. 10 different settings of simulations were done at the same time as the CPU has 12 thread. Total simulation was done in less than 1 minute.

*B. In-situ Conecntration Measurement*

We prepared 100 times more concentrated version of Yamazaki's nutrient solution for lettuce. It consisted of 0.4M of $KNO_3$, 0.1M of $Ca(NO_3)_2$ and 0.05M of $NH_4H_2PO_4$. Chemical compounds which are needed for other ions of

|   | 1 | 2 | 3 | 4 | 5 | 6 | 7 | 8 | 9 | 10 |
|---|---|---|---|---|---|---|---|---|---|---|
| **K** | | | | | | | | | | |
| Theoretical Value | 3.8835 | 7.327 | 10.689 | 13.87 | 16.885 | 19.747 | 22.467 | 25.055 | 27.521 | 29.873 |
| Simulator Prediction | 3.8835 | 7.327 | 10.689 | 13.87 | 16.885 | 19.747 | 22.467 | 25.055 | 27.521 | 29.873 |
| Experimental Value | 4.36955 | 7.96121 | 11.3368 | 14.5715 | 17.55282 | 21.32524 | 24.19406 | 27.28197 | 30.85615 | 34.4981 |
| **Ca** | | | | | | | | | | |
| Theoretical Value | 0.97087 | 1.832 | 2.672 | 3.467 | 4.221 | 4.937 | 5.617 | 6.264 | 6.88 | 7.468 |
| Simulator Prediction | 0.97087 | 1.832 | 2.67199 | 3.46699 | 4.220957 | 4.936942 | 5.616895 | 6.263866 | 6.87983 | 7.4678 |
| Experimental Value | 0.95579 | 1.54849 | 2.04874 | 2.31673 | 2.650174 | 3.119853 | 3.356937 | 3.65157 | 3.790148 | 4.00463 |
| **NO3** | | | | | | | | | | |
| Theoretical Value | 5.82524 | 10.991 | 16.033 | 20.804 | 25.327 | 26.92 | 33.7 | 37.582 | 41.281 | 44.81 |
| Simulator Prediction | 5.82524 | 10.991 | 16.033 | 20.804 | 25.327 | 29.621 | 33.701 | 37.583 | 41.281 | 44.809 |
| Experimental Value | 4.04472 | 8.24084 | 12.1232 | 16.0027 | 19.1123 | 22.95228 | 25.57974 | 29.36491 | 31.64328 | 34.8821 |
| **NH4** | | | | | | | | | | |
| Theoretical Value | 0.48544 | 0.916 | 1.336 | 1.734 | 2.111 | 2.468 | 2.808 | 3.132 | 3.44 | 3.734 |
| Simulator Prediction | 0.48544 | 0.916 | 1.336 | 1.7339 | 2.111 | 2.468 | 2.808 | 3.132 | 3.44 | 3.734 |
| Experimental Value | 0.81084 | 1.46115 | 2.09361 | 2.68363 | 3.104724 | 3.616726 | 4.101231 | 4.388624 | 4.899455 | 5.34784 |

**Table 1. Experiment Results. (mol / mL)**

Yamazaki's nutrient solution were omitted in order to avoid any ions which are unmeasurable with our ISE devices.

By adding the concentrated solution on 1L water, we gradually increased the ionic concentration. Total 10 steps of addition was conducted and theoretical value for individual ions at each experimental step are provided on Table 1.

*C. Simulation*

The network model simulator was designed to receive various parameters: names of components, initial concentration, ionic state, reaction rate coefficient and the stirring velocity of water. We set the initial concentration values of $KNO_3$, $Ca(NO_3)_2$ and $NH_4H_2PO_4$ as the same value from wet experiment's. And we set the initial value of any other components except $H_2O$, $H^+$ and $OH^-$ into 0 in order to make the simulation and experimental condition be same.

The time step dt was set to 1e-8 second to avoid step-update related issue. Although the concentration doesn't show divergence, some ions with low concentration sometimes converged into wrong value when dt was set to 1e-4 second. The authors recommend using smaller time step for each update. Each simulation was terminated after 2.5 million updates, which took less than 50 seconds.

## V. RESULT

The results from in-situ concentration measurement and simulation are provided on Table 1. The values are in mol per milliliter scale.

The simulator predicted theoretical value almost exactly. However, all the experimental value showed significant error. The error becomes greater at higher-concentration condition.

## VI. CONCLUSION

Experiment showed that ion interference effect makes ISE value unclear. The errors of experimental values are not related to calibration or sensor malfunction because the ISE were calibrated with single-ion state solutions, whose concentration is exactly same as the solutions used for wet experiment. As interfering ions disturb Nernst potential on the membrane of ISE, any glass-based sensory device has ion interference issue. Therefore, applying ISEs on industrial condition to maintain nutrient solution is not feasible.

However, the network model provided in this paper has no prediction error even the prediction was gradient-descent based approach rather than calculation of dissociational value of chemical components directly. Applying complex system modeling would help removal of limitation of ISE approach and provide more precise status of nutrient solution system.

ACKNOWLEDGMENT

The authors would like to express our gratitude to Janghun Lee for development of ISE data acquisition software for sensory experiment.

# SUPPLEMENTARY INFORMATIONS

## S1. Equations for simulation

$$\frac{dH_2O}{dt} = -k_0[H_2O] + k_1[H^+][OH^-] \quad (1)$$

$$\frac{dH^+}{dt} = k_0[H_2O] - k_1[H^+][OH^-] \quad (2)$$

$$\frac{dOH^-}{dt} = k_0[H_2O] - k_1[H^+][OH^-] \quad (3)$$

$$\frac{dKNO_3}{dt} = -k_2[KNO_3] + k_3[K^+][NO_3^-] \quad (4)$$

$$\frac{dK^+}{dt} = k_2[KNO_3] - k_3[K^+][NO_3^-] \quad (5)$$

$$\frac{dNO_3^-}{dt} = k_2[KNO_3] - k_3[K^+][NO_3^-] \quad (6)$$

$$\frac{dCa(NO_3)_2}{dt} = -k_4[Ca(NO_3)_2] + k_5[Ca^{2+}][NO_3^-]^2 \quad (7)$$

$$\frac{dCa^{2+}}{dt} = k_4[Ca(NO_3)_2] - k_5[K^+][NO_3^-]^2 \quad (8)$$

$$\frac{dNO_3^-}{dt} = 2k_4[Ca(NO_3)_2] - k_5[K^+][NO_3^-]^2 \quad (9)$$

$$\frac{dNH_4H_2PO_4}{dt} = -k_6[NH_4H_2PO_4] + k_7[NH_4^+][H_2PO_4^-] \quad (10)$$

$$\frac{dNH_4^+}{dt} = k_6[NH_4H_2PO_4] - k_7[NH_4^+][H_2PO_4^-] \quad (11)$$

$$\frac{dH_2PO_4^-}{dt} = k_6[NH_4H_2PO_4] - k_7[NH_4^+][H_2PO_4^-] \quad (12)$$

$$\frac{dH_3PO_4}{dt} = -k_8[H_3PO_4] + k_9[H^+][H_2PO_4^-] \quad (13)$$

$$\frac{dH^+}{dt} = k_8[H_3PO_4] - k_9[H^+][H_2PO_4^-] \quad (14)$$

$$\frac{dH_2PO_4^-}{dt} = k_8[H_3PO_4] - k_9[H^+][H_2PO_4^-] \quad (15)$$

$$\frac{dH_2PO_4^-}{dt} = -k_{10}[H_2PO_4^-] + k_{11}[H^+][HPO_4^{2-}] \quad (16)$$

$$\frac{dH^+}{dt} = k_{10}[H_2PO_4^-] - k_{11}[H^+][HPO_4^{2-}] \quad (17)$$

$$\frac{dHPO_4^{2-}}{dt} = k_{10}[H_2PO_4^-] - k_{11}[H^+][HPO_4^{2-}] \quad (18)$$

$$\frac{dHPO_4^{2-}}{dt} = -k_{12}[HPO_4^{2-}] + k_{13}[H^+][PO_4^{3-}] \quad (19)$$

$$\frac{dH^+}{dt} = k_{12}[HPO_4^{2-}] - k_{13}[H^+][PO_4^{3-}] \quad (20)$$

$$\frac{dPO_4^{3-}}{dt} = k_{12}[HPO_4^{2-}] - k_{13}[H^+][PO_4^{3-}] \quad (21)$$

$$\frac{dCa^{2+}}{dt} = k_{14}[CaHPO_4] - k_{15}[Ca^{2+}][HPO_4^{2-}] \quad (22)$$

$$\frac{dHPO_4^{2-}}{dt} = k_{14}[CaHPO_4] - k_{15}[Ca^{2+}][HPO_4^{2-}] \quad (23)$$

$$\frac{dCaHPO_4}{dt} = -k_{14}[CaHPO_4] + k_{15}[Ca^{2+}][HPO_4^{2-}] \quad (24)$$

$$\frac{dCa^{2+}}{dt} = 3k_{16}[Ca_3(PO_4)_2] - k_{17}[Ca^{2+}]^3[PO_4^{3-}]^2 \quad (25)$$

$$\frac{dHPO_4^{2-}}{dt} = 2k_{16}[Ca_3(PO_4)_2] - k_{17}[Ca^{2+}]^3[PO_4^{3-}]^2 \quad (26)$$

$$\frac{dCa(PO_4)_2}{dt} = -k_{16}[Ca_3(PO_4)_2] + k_{17}[Ca^{2+}]^3[PO_4^{3-}]^2 \quad (27)$$

$$\frac{dCa^{2+}}{dt} = k_{18}[Ca(H_2PO_4)_2] - k_{19}[Ca^{2+}][H_2PO_4^-]^2 \quad (28)$$

$$\frac{dH_2PO_4^-}{dt} = 2k_{18}[Ca(H_2PO_4)_2] - k_{19}[Ca^{2+}][H_2PO_4^-]^2 \quad (29)$$

$$\frac{dCa(H_2PO_4)_2}{dt} = -k_{18}[Ca(H_2PO_4)_2] + k_{19}[Ca^{2+}][H_2PO_4^-]^2 \quad (30)$$

## S2. $K_{sp}$ and $K_a$

(1) $H_2O \leftrightarrow H^+ + OH^-$

$K_a = [OH^-]^2 = (10^{-7})^2 = 10^{-14}$ at pH 7

(2) $KNO_3 \leftrightarrow K^+ + NO_3^-$

$K_{sp} = [K^+][NO_3^-] = (3.77685133)^2[24] = 14.26460597$

(3) $Ca(NO_3)_2 \leftrightarrow Ca^{2+} + 2NO_3^-$

$K_{sp} = [Ca^{2+}][NO_3^-]^2 = 8.7495048(2 \times 8.7495048)^2[24] = 2679.232594$

(4) $NH_4H_2PO_4 \leftrightarrow NH_4^+ + H_2PO_4^-$

$K_{sp} = [NH_4^+][H_2PO_4^-] = (3.5017430)^2[24] = 12.262204$

(5) $H_3PO_4 \leftrightarrow H^+ + H_2PO_4^-$, $K_a = 0.00707946$ [25]

(6) $H_2PO_4^- \leftrightarrow H^+ + HPO_4^{2-}$, $K_a = 8.1283\text{e-}08$ [25]

(7) $HPO_4^{2-} \leftrightarrow H^+ + PO_4^{3-}$, $K_a = 4.7863\text{e-}13$ [25]

(8) $CaHPO_4 \leftrightarrow Ca^{2+} + HPO_4^{2-}$

$K_{sp} = [Ca^{2+}][HPO_4^{2-}] = (0.0014655)^2[24] = 2.14787e - 6$

(9) $Ca_3(PO_4)_2 \leftrightarrow 3Ca^{2+} + 2PO_4^{3-}$

$K_{sp} = [Ca^{2+}]^3[PO_4^{3-}]^2$
$= (2 \times 3.857e - 6)^2(3 \times 3.857e - 6)^3[24] = 9.22e - 26$

(10) $Ca(H_2PO_4)_2 \leftrightarrow Ca^{2+} + 2H_2PO_4^-$

$K_{sp} = [Ca^{2+}][H_2PO_4^-]^2 = (0.0769)(2 \times 0.0769)^2[26]$
$= 0.00591361$

## S3. $k_f$ and $k_b$

| Coefficient | Value | Coefficient | Value |
|---|---|---|---|
| $k_0$ | 1e-20 | $k_1$ | 1e-6 |
| $k_2$ | 976.8870716 | $k_3$ | 0 |
| $k_4$ | 161.1897361 | $k_5$ | 0 |
| $k_6$ | 105.7203812 | $k_7$ | 0 |
| $k_8$ | 0.725 | $k_9$ | 100 |
| $k_{10}$ | 6.31e-6 | $k_{11}$ | 100 |
| $k_{12}$ | 3.98e-13 | $k_{13}$ | 100 |
| $k_{14}$ | 1e-5 | $k_{15}$ | 100 |
| $k_{16}$ | 1.2e-16 | $k_{17}$ | 100 |
| $k_{18}$ | 0.591361 | $k_{19}$ | 100 |

Table S1. $k_f$ and $k_b$ values for simulation